\documentclass[runningheads]{llncs}
\usepackage{multirow}
\usepackage{graphicx}
\usepackage{amsmath}

\graphicspath{ {figures/}}
\usepackage{comment}
\usepackage{hyperref}
\hypersetup{
    colorlinks=true,
    linkcolor=blue
}
\usepackage{bbding}
\usepackage{pifont}
\usepackage{wasysym}
\usepackage{amssymb}
\usepackage{booktabs}

\usepackage{biblatex}
\usepackage{soul}
\usepackage[dvipsnames]{xcolor}
\makeatletter
\newcommand{\@chapapp}{\relax}%
\makeatother
\newcommand{\cmark}{\ding{51}}

\usepackage[title,toc,titletoc,header]{appendix}

\begin{document}

\title{Cohort Bias Adaptation in Aggregated Datasets for Lesion Segmentation}

\author{Brennan Nichyporuk \inst{1,2} \and
Jillian Cardinell \inst{1,2} \and
Justin Szeto \inst{1,2} \and
Raghav Mehta \inst{1,2} \and
Sotirios Tsaftaris \inst{3,4} \and
Douglas L. Arnold \inst{5,6} \and
Tal Arbel \inst{1,2}
}


\authorrunning{B. Nichyporuk \textit{et al.}}

\institute{Centre for Intelligent Machines, McGill University, Canada 
\and 
MILA  (Quebec  Artificial  Intelligence  Institute),  Montreal, Canada
\and
Institute for Digital Communications, School of Engineering, University of Edinburgh, UK
\and
The Alan Turing Institute, UK
\and
Department of Neurology and Neurosurgery, McGill University, Canada
\and 
NeuroRx Research, Montreal, Canada \\
\email{brennann@cim.mcgill.ca}}

\maketitle              
\begin{abstract}

Many automatic machine learning models developed for focal pathology (e.g. lesions, tumours) detection and segmentation perform well, but do not generalize as well to new patient cohorts, impeding their widespread adoption into real clinical contexts. One strategy to create a more diverse, generalizable training set is to naively pool datasets from different cohorts. Surprisingly, training on this {\it big data} does not necessarily increase, and may even reduce, overall performance and model generalizability, due to the existence of cohort biases that affect label distributions. In this paper, we propose a generalized affine conditioning framework to learn and account for cohort biases across multi-source datasets, which we call Source-Conditioned Instance Normalization (SCIN). Through extensive experimentation on three different, large scale, multi-scanner, multi-centre Multiple Sclerosis (MS) clinical trial MRI datasets, we show that our cohort bias adaptation method (1) improves performance of the network on pooled datasets relative to naively pooling datasets and (2) can quickly adapt to a new cohort by fine-tuning the instance normalization parameters, thus learning the new cohort bias with only 10 labelled samples.

\keywords{deep learning \and multiple sclerosis  \and instance norm conditioning \and trial merging \and segmentation \and label style \and cohort bias.}
\end{abstract}

\section{Introduction}
A large number of deep learning (DL) models have been developed and successfully applied to the contexts of healthy structure and pathology segmentation (e.g. brain tumours), with good performance on a number of public datasets \cite{ronneberger2015u, kamnitsas2017efficient}. However, DL methods are generally not able to overcome large mismatches between the training and testing distributions. Several recent papers have focused on addressing poor performance and generalizability resulting from differences in imaging acquisition parameters, resolutions, and scanners differences \cite{biberacher_intra-_2016, karani_lifelong_2018, van_opbroek_transfer_2015}, missing modalities and sequences \cite{HeMIS, shen_brain_2019}, or overcoming labelling style differences between individual raters~\cite{chotzoglou_exploring_2019, vincent_impact_2021, jungo_effect_2018, shwartzman_worrisome_2019}. 

In the context of focal pathology (e.g. lesions, tumours) detection and segmentation, the inability to generalize methods to new patient cohorts seriously impedes their widespread adoption into real clinical contexts. However, the problem of generalization is much more subtle in this context, particularly given the lack of real "ground truth" labels. Even in cohorts with similar image acquisition parameters and labelling protocols, significant biases will still exist due to differences between patient populations. Indeed, even if the same set of raters label both cohorts, the probability of labeling any given voxel can still depend on rater knowledge about the overall distribution of each patient population. One obvious strategy to create a more diverse, generalizable training set is to naively pool datasets from different cohorts. However, this will not necessarily increase, and may even decrease, overall performance and model generalizability due to biases that ultimately affect the label distribution of each cohort.

In this work, we propose a generalized conditioning framework to learn and account for cohort biases, and associated annotation {\it style}, across multi-source pooled datasets. The approach entails the use of conditional instance normalization layers \cite{CIN} to condition the network on auxiliary cohort information in an approach we call Source-Conditioned Instance Normalization (SCIN). This allows the network to leverage a pooled dataset while providing cohort-specific segmentations. A key advantage of the proposed approach is that it can be adapted to a new clinical dataset if provided with a small subset of labelled samples. Experiments are performed on three different large-scale, multi-scanner, multi-center multiple sclerosis (MS) clinical trial datasets. In particular, we show that our cohort bias adaptation method can (1) learn an annotation protocol specific to each cohort that makes up aggregated dataset without loss in performance, (2) improves performance of the network on pooled datasets relative to naively pooling datasets, and (3) can quickly adapt an already conditioned DL model on a new clinical trial dataset with only 10 labeled samples, by fine-tuning the conditional instance normalization parameters. Finally, to illustrate that our model is accounting for complex cohort biases, we artificially create a sub-trial cohort without small lesions labelled, and show that our model is able to learn this cohort bias and account for it. 

\section{Related work}
\label{sec:related}
Several papers have focused on accounting for the differences in image acquisition parameters or missing modalities \cite{biberacher_intra-_2016, karani_lifelong_2018, van_opbroek_transfer_2015}. However, in clinical trials, these variables are usually kept constant or are vigorously normalized. On the other hand, several papers have examined label bias, focusing on inter-rater bias or modelling an individual rater's label style \cite{heller_imperfect_2018, warfield_simultaneous_2004, joskowicz_inter-observer_2019, kohl_hierarchical_2019}. In these cases, each rater is provided the same cohort to label, and inter-rater biases can be attributed to differences in experience level or lesion boundary uncertainty \cite{jungo_effect_2018}. Furthermore, in many cases a simplifying assumption, that each rater is a noisy reflection of ground truth, is made \cite{ji_learning_2021, zhang_disentangling_2020}. Despite these shortcomings, inter-rater studies have stressed the importance of considering biases that can affect labels, with Schwartzman \textit{et al} \cite{shwartzman_worrisome_2019} showing that rater bias in training samples is actually amplified by neural networks. Other researchers have also found that multiple labels with varying biases can provide important insights into uncertainty estimation \cite{chotzoglou_exploring_2019, vincent_impact_2021}.

\section{Methods}
\label{sec:methods}
\begin{figure}[t]
\centering
  \centering
   \includegraphics[width=0.95\textwidth]{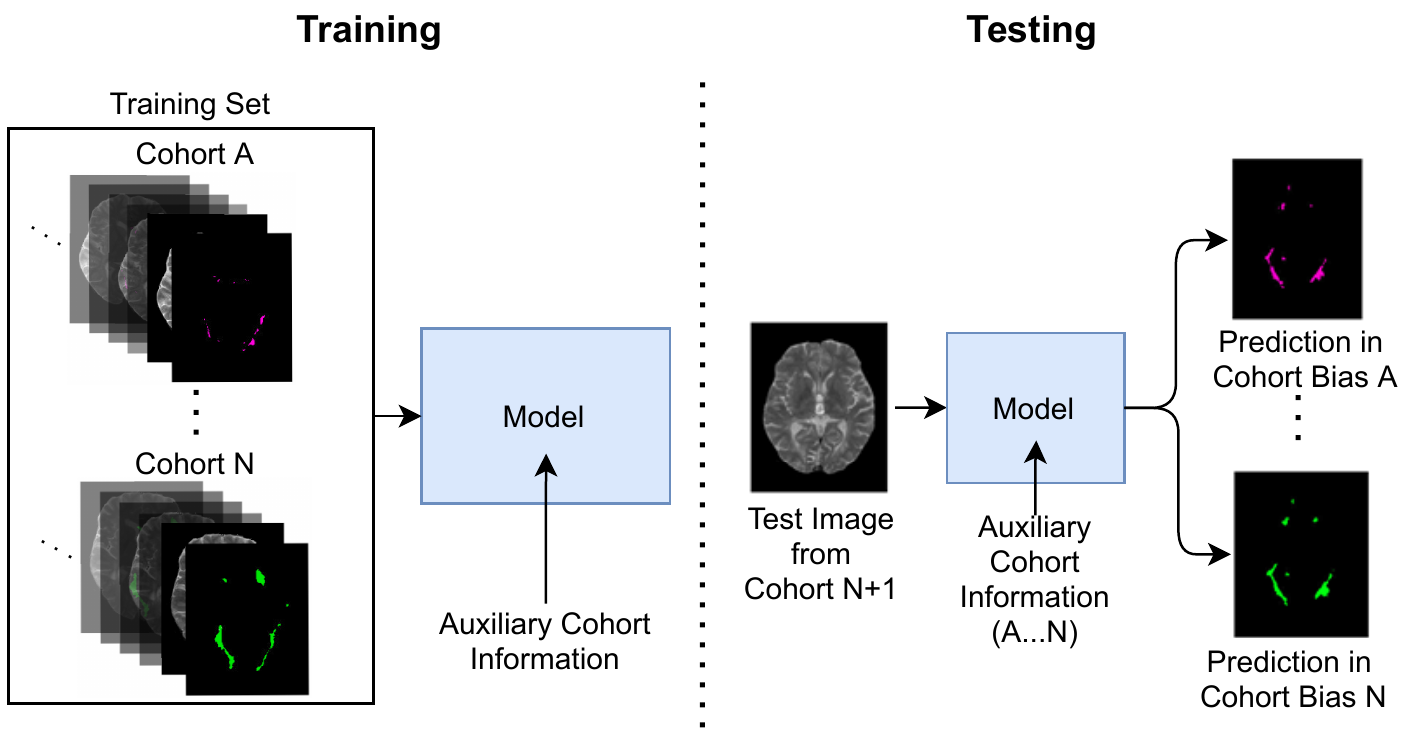}
  \caption{System overview showing training on the left and testing on the right. The left shows how we train with multiple cohorts and use auxiliary cohort information to learn the associated bias. On the right is how we use cohort information during testing to generate multiple labels for an image in a desired style.}
  \label{fig:flow}
\end{figure}

We propose a framework that is able to learn cohort-specific biases which are ultimately manifested in the 'ground truth' labels. To do this, we employ the conditional instance normalization mechanism proposed by Vincent \textit{et al} \cite{CIN} to model cohort biases with source-specific parameters. In this paper, we refer to the whole approach as Source-conditioned Instance Normalization (SCIN), and the conditioning mechanism proposed by Vincent et. al. as Conditional Instance Normalization (CIN) \cite{CIN}. In practice, this involves scaling/shifting the normalized activations at each layer by using the set of affine parameters corresponding to a given sample's source cohort during the forward pass. As a result, source-specific affine parameters are learned only from samples that make up the cohort itself.  Mathematically, the mechanism is represented by the following equation:

$$\text{CIN} (z) = \gamma_s \left(\frac{z-\mu(z)}{\sigma(z)}\right) + \beta_s$$
where $\gamma_s$ and $\beta_s$ are the affine parameters specific to cohort source $s$, and where $\mu(z)$ and $\sigma(z)$ represent the per-channel mean and standard deviation, respectively. Other than cohort-specific parameters, all other network parameters are learned from all samples regardless of cohort identity. This allows the approach to pool multiple datasets, leveraging the aggregated dataset to learn common features while still taking into account cohort-specific biases. A full system overview can be found in Figure \ref{fig:flow}.

\section{Implementation Details}
\subsection{Network Architecture and Training Parameters}

\begin{figure}[t]
\centering
  \centering
   \includegraphics[width=\textwidth]{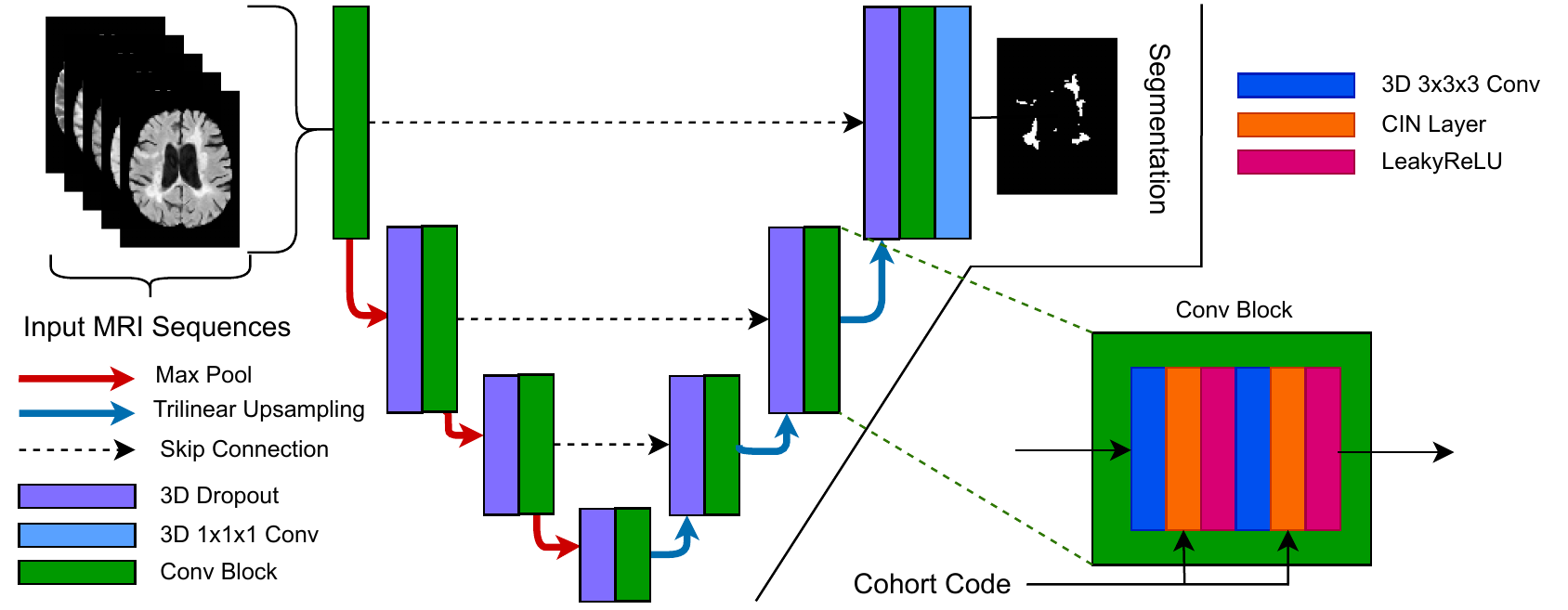}
  \caption{Left: Overview of modified nnUNet \cite{nnUnet} architecture used to segment MS T2 lesions. Right: Detail of a conv block. It consists of a series of 3D 3x3x3 Convolution Layer, CIN layer, and a LeakyReLU activation layer.}
  \label{fig:architecture}
\end{figure}

We utilize a modified version of nnUNet~\cite{nnUnet} for MS T2-weighted lesion segmentation. Specifically, we add dropout and CIN as depicted in Figure \ref{fig:architecture}. All models in this paper are trained using Binary Cross Entropy (BCE).

\subsection{Data Set}
We make use of a large proprietary MS dataset from three different clinical trials, where each trial contains multi-modal MR sequences of patients with different disease subtypes, and/or at different stages of disease. Specifically, Trial-A is a late-stage Secondary-Progressive (SPMS) dataset with 1000 patients, Trial-B is a Relapsing Remitting (RRMS) dataset with 1000 patients, and Trial-C is an early-stage SPMS dataset with 500 patients. Each patient sample consists of 5 MR sequences (T1-weighted, T1-weighted with gadolinium contrast agent, T2-weighted, Fluid Attenuated Inverse Recovery, and Proton Density). All MR sequences were acquired at 1mm x 1mm x 3mm resolution. T2 lesion labels were generated at the end of the clinical trial, and were produced through an external process where trained expert annotators manually corrected a proprietary automated segmentation method. All data was processed using similar processing pipeline(s), but at different points in time (i.e. the clinical trials were not all held at the same time) and with different versions of the baseline segmentation method. Although different expert raters corrected the labels, there was overlap between raters across trials and all raters were trained to follow a similar labelling protocol. All three trial datasets are divided into non-overlapping training (60\%), validation (20\%), and testing (20\%) sets. 

\subsection{Evaluation Metrics}
Automatic T2 lesion segmentation methods are evaluated with voxel-level segmentation, and for the small lesion removal experiments in section \ref{sec:msl}, lesion-level detection metrics. Specifically, we use DICE score \cite{F1} for voxel-level segmentation and F1-score for lesion-level detection analysis \cite{commowick_objective_2018}. \footnote{Operating point (threshold) is chosen based on the best voxel-level DICE} All methods utilized in this work output a voxel-level segmentation mask. A connected component analysis is performed on the voxel-based segmentation mask to group lesion voxels in an 18-connected neighbourhood \cite{BUNet}. Given that MS lesions vary significantly in size, we report results stratified by lesion size.

\section{Experiments and Results}
We perform three different sets of experiments to demonstrate the usefulness of the proposed SCIN approach. In the first experiment, we show that SCIN is able to strategically pool diverse datasets with differing cohort biases. The second experiment demonstrates the clinical utility of SCIN to adapt to new cohort biases with limited available labeled data. Finally, we show that SCIN is able to model complex cohort biases by simulating a type of cohort bias where small lesions (10 voxels or less) were not labeled.

\subsection{Trial Conditioning}
\label{Sec:Experiment-1}
Experiments in this section aim to show how the proposed SCIN approach allows for pooling of data from multiple cohorts while taking into account cohort specific biases. We use two different clinical trial datasets (Trial-A and Trial-B) for these experiments. These two trials were collected several years apart with patients of different disease subtypes. Given that each trial is processed independently and at different points in time, minor differences in site/scanner distribution and annotation style will exist. Together, the patient population, site/scanner distribution, and annotation style create a distinct cohort bias, which we aim to account for with the proposed method.

We train four different models on these datasets: (i) a model trained on only Trial-A, with Instance Norm (IN)~\cite{ulyanov_instance_2016}, (ii) a model trained on only Trial-B (with IN), (iii) a model trained on a naively pooled dataset consisting of Trial-A and Trial-B (with IN), and (iv) a model trained on both Trial-A and Trial-B using the SCIN approach. All four models were tested on the same held-out test set from both trials.

\begin{table}[t]
\scriptsize
\centering
\caption{Dice scores shown on Trial-A and Trial-B test sets for models trained with different combinations of Trial-A and Trial-B training sets. Trial-A and Trial-B training sets each contain 600 patients.} \label{tab:AB-metrics}
\begin{tabular}{|c|c|cc|cc|cc|}
\hline
\multirow{2}{*}{} & \multirow{2}{*}{\textbf{Model}}       & \multicolumn{2}{c|}{\textbf{Train Set}}                         & \multicolumn{2}{c|}{\textbf{Conditioned On}}             & \multicolumn{2}{c|}{\textbf{Test Performance}}           \\ \cline{3-8} 
                  &                                       & \multicolumn{1}{c|}{\textbf{Trial-A}} & \textbf{Trial-B}        & \multicolumn{1}{c|}{\textbf{Trial-A}} & \textbf{Trial-B} & \multicolumn{1}{c|}{\textbf{Trial-A}} & \textbf{Trial-B} \\ \hline
1                 & \textbf{Single-Trial}                    & \cmark                                &                         & \multicolumn{2}{c|}{-}                                   & 0.793                                 & 0.689            \\
2                 & \textbf{Single-Trial}                    &                                       & \cmark                  & \multicolumn{2}{c|}{-}                                   & 0.715                                 & 0.803            \\
3                 & \textbf{Naive-Pooling}                    & \cmark                                & \cmark                  & \multicolumn{2}{c|}{-}                                   & 0.789                                 & 0.748            \\ \hline
4                 & \multirow{2}{*}{\textbf{SCIN-Pooling}} & \multirow{2}{*}{\cmark}               & \multirow{2}{*}{\cmark} & \cmark                                &                  & 0.794                                 & 0.700            \\
5                 &                                       &                                       &                         &                                       & \cmark           & 0.725                                 & 0.797            \\ \hline
\end{tabular}
\end{table}

\begin{figure}[t]
\centering
  \centering
   \includegraphics[width=\textwidth]{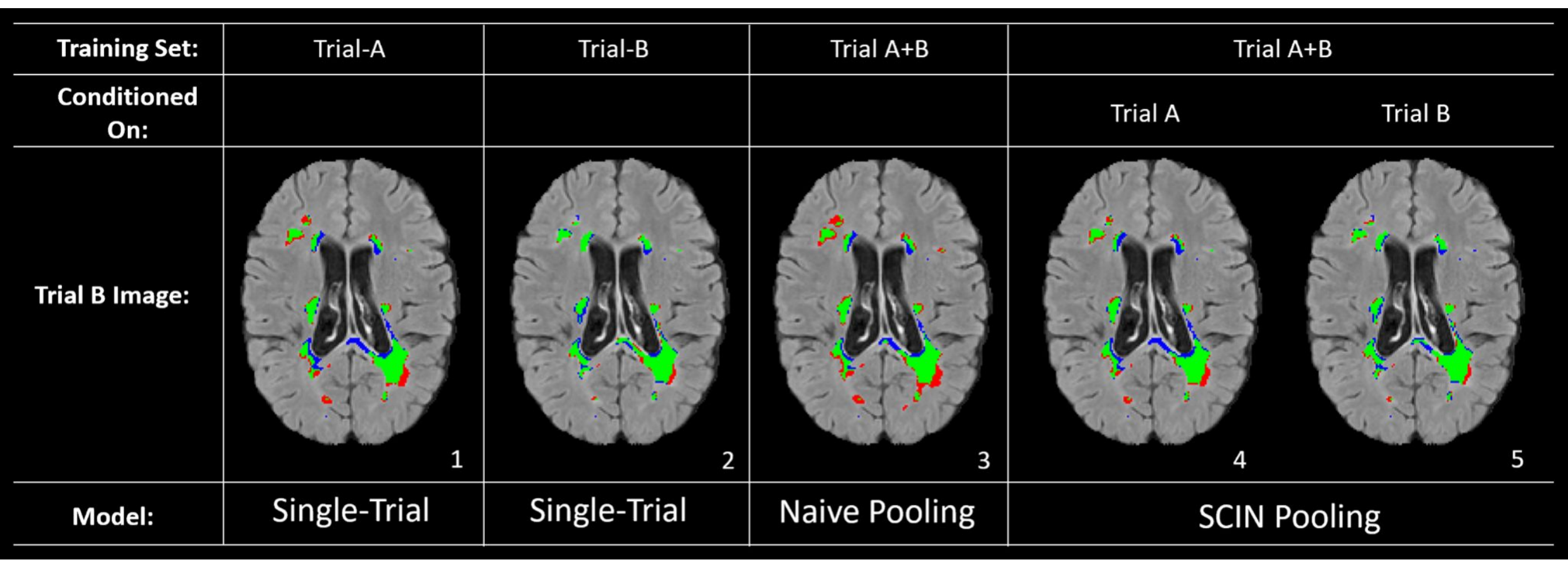}
  \caption{Qualitative lesion segmentation labels (red is false positives, blue is false negatives, green is true positives) superimposed onto a FLAIR test image from Trial B. The results are based on the models from Rows 1-5 (left to right) of Table 1.} 
  \label{fig:vis}
\end{figure}

Table \ref{tab:AB-metrics} depicts the performance of the aforementioned models on the hold-out test sets. Results indicate that models trained on only one trial (Row-1 and Row-2) generalize poorly when tested on the other trial. A model trained on the naively pooled dataset, consisting of data from both trials (Row-3), shows better generalization across trials, but still falls short of the performance achieved by each trial-specific model, especially in the case of the Trial-B dataset. At first glance, this might appear surprising given the expectation that a model trained on a larger, pooled dataset should generally perform better relative to a model that has access to less data. However, given the knowledge that biases can exist between cohorts, it is no mystery why the naively pooled model would be unable to generate a labeling consistent with the cohort of the sample, given that the cohort / labeling bias cannot be identified from the image alone. On the other hand, a single SCIN-pooling model conditioned on each trial (Row-4 and Row-5), is able to learn trial-specific parameters to model the bias specific to the trial in question, improving performance relative to the naively-pooled model (Row-3). Note that using the incorrect set of trial-specific parameters with the SCIN-Pooling model (Row-4 and Row-5) results in a performance decline similar to that observed when testing the Single-Trial models (Row-1 and Row-2) on the corresponding unseen trial. This simply serves as a sanity check, and confirms that the proposed method effectively models the cohort bias of each dataset.

Qualitative results for labels produced by different models on a single Trial-B test case are shown in Figure \ref{fig:vis}. From this, we can see that generating labels on a Trial-B test case using a Single-Trial model trained on the Trial-A dataset (Image-1) leads to an increased number of false positive and false negative voxels. This is also true when testing a naively pooled model (Image-3). On the other hand, the proposed SCIN-pooled model conditioned on Trial-B (Image 5) does not suffer from a significant degradation in segmentation quality, showing that the SCIN approach enables leveraging multiple datasets with different cohort biases without significant performance decrements. Visually, note that the labeling style of the the SCIN-pooled model is similar to that of the corresponding Single-Trial model, showing that SCIN-pooled method is able to learn a cohort-specific bias for each trial.  

\begin{table}[t]
\scriptsize
\centering
\caption{Dice scores shown on the Trial-C test set from the Naive-pooling and SCIN-pooling models trained on Trial A and B. Dice scores are also shown for fine-tuned versions of those models, where the IN parameters were tuned using 10 Trial-C samples.} 
\label{tab:finetune-metrics}
\begin{tabular}{|c|c|c|ccc|c|}
\hline
\multirow{2}{*}{} & \multirow{2}{*}{\textbf{Model}}       & \multirow{2}{*}{\textbf{\begin{tabular}[c]{@{}c@{}}Fine-Tuned\\ on Trial-C\end{tabular}}} & \multicolumn{3}{c|}{\textbf{Conditioned on}}                                                     & \multirow{2}{*}{\textbf{Test Performance}} \\ \cline{4-6}
                  &                                       &                                                                                           & \multicolumn{1}{c|}{\textbf{Trial-A}} & \multicolumn{1}{c|}{\textbf{Trial-B}} & \textbf{Trial-C} &                                            \\ \hline
1                 & \multirow{2}{*}{\textbf{Naive-Pooling}}   &                                                                                           & \multicolumn{3}{c|}{-}                                                                           & 0.774                                      \\
2                 &                                       & \cmark                                                                                    & \multicolumn{3}{c|}{-}                                                                           & 0.819                                      \\ \hline
3                 & \multirow{3}{*}{\textbf{SCIN-Pooling}} &                                                                                           & \cmark                                &                                       &                  & 0.763                                      \\
4                 &                                       &                                                                                           &                                       & \cmark                                &                  & 0.806                                      \\
5                 &                                       & \cmark                                                                                    &                                       &                                       & \cmark           & 0.834                                      \\ \hline
\end{tabular}
\end{table}

\subsection{Fine-tuning to New Cohort Bias}
The second set of experiments aims to mimic a clinical situation where large datasets (Trial-A and Trial-B) have a known cohort bias. A new small dataset (Trial-C) is provided with an unidentified cohort bias. We take two pre-trained models (Naively-Pooled and SCIN-pooled) from the previous set of experiments (see Sec \ref{Sec:Experiment-1}), and fine-tune the affine parameters of the IN/CIN layers with only 10 labeled samples from the Trial-C dataset. Segmentations are performed on a hold-out test set from Trial-C. Similar to Experiment 1, time of collection and disease subtype were the primary differences between the three trials (along with minor differences in scanner/site distribution and labeling protocol).

Table \ref{tab:finetune-metrics} depicts the results for this set of experiments. We can see that the performance of the naively-pooled model improves when the IN parameters of the model are fine-tuned (Row-2) compared to no fine-tuning (Row-1). Furthermore, a SCIN-pooled model shows good Trial-C performance when conditioned on Trial-B (Row-4), indicating that Trial-C has similar cohort biases. By fine-tuning the trial-specific CIN layer parameters of this model on 10 samples of Trial-C, we are able to then condition the model on Trial-C during test time. This leads to the highest performance improvement (Row-5) over all models, including fine-tuning the naively pooled model (Compare Row-2 and Row-5). This shows that with SCIN, we can more effectively learn features common to both Trail-A and Trial-B, resulting in better performance after fine-tuning on a hold-out trial.

\subsection{Accounting for Complex Cohort Biases - Missing Small Lesions}
\label{sec:msl}
The final set of experiments examine whether the SCIN approach is able to learn complex non-linear cohort biases. Accordingly, we isolate biases that arise solely from different labelling protocols while keep all other factors, such as disease stage and time of collection, constant. To that end, we utilize a held-out clinical trial dataset (Trial-C) and artificially modify half of the dataset by removing small lesions (10 voxels or less) from the provided labels. This can be thought of as being equivalent to a labeling protocol that misses or ignores small lesions (Trial-MSL). The labels of the remaining half of the dataset are not modified in any way (Trial-Orig). 

\begin{table}[t]
\scriptsize
\centering
\caption{Voxel based Dice scores and small lesion detection F1 scores shown on Trial-C (Trial-Orig) held-out test set using models trained on different combinations of the original dataset (Trial-Orig, 150 training patients) and the dataset with missing small lesions (Trial-MSL, 150 training patients)}
\begin{tabular}{|c|c|cc|cc|cc|}
\hline
\multirow{2}{*}{} & \multirow{2}{*}{\textbf{Model}}       & \multicolumn{2}{c|}{\textbf{Train Set}}                            & \multicolumn{2}{c|}{\textbf{Conditioned On}}                  & \multicolumn{2}{c|}{\textbf{Test Performance}}                      \\ \cline{3-8} 
                  &                                       & \multicolumn{1}{c|}{\textbf{Trial-Orig}} & \textbf{Trial-MSL}      & \multicolumn{1}{c|}{\textbf{Trial-Orig}} & \textbf{Trial-MSL} & \multicolumn{1}{c|}{\textbf{Sm Lesion F1}} & \textbf{Voxel Dice} \\ \hline
1                 & \textbf{Single-Trial}                    & \cmark                                   &                         & \multicolumn{2}{c|}{-}                                        & 0.795                                         & 0.844               \\
2                 & \textbf{Single-Trial}                    &                                          & \cmark                  & \multicolumn{2}{c|}{-}                                        & 0.419                                         & 0.837               \\
3                 & \textbf{Naive-Pooling}                    & \cmark                                   & \cmark                  & \multicolumn{2}{c|}{-}                                        & 0.790                                         & 0.797               \\ \hline
4                 & \multirow{2}{*}{\textbf{SCIN-Pooling}} & \multirow{2}{*}{\cmark}                  & \multirow{2}{*}{\cmark} & \cmark                                   &                    & 0.784                                         & 0.854               \\
5                 &                                       &                                          &                         &                                          & \cmark             & 0.496                                         & 0.850               \\ \hline
\end{tabular}
\label{tab:m3-MSL-metrics}
\end{table}

Table \ref{tab:m3-MSL-metrics} shows the results for this set of experiments on a non-modified Trial-C test set (Trial-Orig). We report detailed results specific to the detection of small lesions in order to examine whether the proposed strategy learns to account for a labeling style that ignores small lesions. The results in Row-1 show that when both train set and test set have the same labeling protocol (Trial-Orig), the Single-Trial model performs well according to both lesion-level detection and voxel-level segmentation metrics. On the other hand, when there is a significant shift in the labeling protocol between the train and test set, the Single-Trial model trained on the Trial-MSL dataset (Row-2) exhibits poor small lesion detection performance. The degradation in small lesion detection performance is expected as Trial-MSL has small lesions labeled as background, while the test set (Trial-Orig) has small lesions marked as lesions. The Naively-Pooled model (Row-3), which is trained on both Trial-Orig and Trial-MSL, learns to completely ignore the bias of the Trial-MSL dataset as measured by lesion-level detection performance. However, voxel-level segmentation performance suffers significantly. On the other hand, a model trained using the SCIN approach is able to adapt to the difference in labeling styles and exhibit good lesion-level detection and voxel-level segmentation performance when conditioned on the appropriate cohort (Row-4). Looking at SCIN-Pooling conditioned on Trial-MSL (Row-5), we see that SCIN is able to learn the Trial-MSL label bias quite effectively, and is able to ignore small lesions while maintaining voxel-level segmentation performance. This shows that SCIN is able to model complex non-linear labeling biases -- its not just a matter of over or under segmentation.

\section{Conclusions}
In this paper, we proposed SCIN, an approach that learns source-specific IN parameters, effectively modeling the bias of each cohort present in an aggregated dataset. We show that the IN parameters of a pre-trained SCIN model can be fine-tuned, allowing the model to learn the bias of an independent cohort with very little data. We demonstrate that the biases learned can be non-linear, resulting in complex differences in the segmentation outputs corresponding to each cohort. Most importantly, we show that proposed method makes it possible to train a high performance model on an aggregate dataset, avoiding the performance penalty observed with naive pooling. Overall, SCIN is simple to implement, and can potentially benefit any application that wishes to leverage a large aggregated dataset in the context of segmentation.

\section*{Acknowledgement}
The authors are grateful to the International Progressive MS Alliance for supporting this work (grant number: PA-1412-02420), and to the companies who generously provided the clinical trial data that made it possible: Biogen, BioMS, MedDay, Novartis, Roche / Genentech, and Teva. Funding was also provided by the Canadian Institute for Advanced Research (CIFAR) Artificial Intelligence Chairs program. S.A. Tsaftaris acknowledges the support of Canon Medical and the Royal Academy of Engineering and the Research Chairs and Senior Research Fellowships scheme (grant RCSRF1819\textbackslash8\textbackslash25).

\printbibliography
\end{document}